# Sintering-coalescence transition on the nanoscale as a bifurcation phenomenon: molecular dynamics study

V.V. Puytov, I.V. Talyzin and V.M. Samsonov[1]
Tver State University, Zhelyabova 33, Tver, 170100, Russia

E-mail: samsonoff@inbox.ru, samsonov.VM@tversu.ru

**Abstract.** Using the isothermal molecular dynamics (MD), coalescence/sintering of Au nanoparticles (NPs) was simulated by employing the Nose-Hoover thermostat. The MD simulation was realized by using the well-known open program LAMMPS, its version for parallel calculations on GPUs. We have found that the solid NP sintering scenario is switched to the coalescence scenario not at the NP melting temperature $T_\mathrm{m}$ exactly but at a lower temperature $T_0 \approx 0.9 T_\mathrm{m}$ interpreted as the critical temperature corresponding to a coalescence/sintering bifurcation phenomenon: in the temperature range from $T_0 - 2$ K to $T_0 + 2$ K the resulting (daughter) NPs of the same size can have either liquid-like or crystalline structure after coalescence/sintering at the same fixed temperature. The crystallize and liquid states were identify by analyzing the degree of crystallinity and the radial distribution function. For this purpose we employed the OVITO program.

## 1. Introduction

The present paper may be considered as at least twice extended version of our short paper "Bifurcation phenomenon in molecular dynamics model of coalescence/sintering on the nanoscale" accepted for publication by "Journal of Physics Conference Series" (open access). This short communication was prepared after an oral presentation at International conference "Chemical Thermodynamics and Kinetics" (G. Novgorod, Russia, May 2021). We have added a number of figures reflecting our recent molecular dynamics (MD) results, not presented in our previous papers.

The term 'coalescence' is usually referred to merging of two droplets, i.e. formation of a larger daughter one whereas the term 'sintering' relates rather to agglomeration of solid particles without their complete merging. Since 90-s a great attention has been paid to coalescence and sintering processes on the nanoscale. On the one hand, sintering of nanoparticles (NPs) is used in powder metallurgy to produce some new nanostructured materials, in additive manufacturing and other directions of nanotechnology. On the other hand, the complete merging of NPs, i.e. coalescence as well as the grain coarsening can be undesirable when a nanostructured material is manufactured with some controlled grain size or when metal powders are used as catalysts. Really, most of catalysts are used in high-temperature conditions when without any protective additives fine NPs tend to agglomerate. An analogous situation may occur in flexible cooling devices and other devices on the basis of metal NPs.

---
[1] To Whom correspondence should be addressed

No doubt that regularities and mechanisms of the coalescence and sintering processes should depend on temperature and, in particular, on the NP melting temperature $T_\mathrm{m}$ which, in turn, depends on the NP size, i.e. its radius $r_0$ or the number of atoms $N_0$ the NP consists of. So, in most papers, devoted to MD simulation of the coalescence and sintering processes, the size dependence of $T_\mathrm{m}$ is also evaluated [1-6]. However, the relationship between the NP melting temperature and coalescence/sintering of NPs of the chosen size does not seem to be quite definitely formulated and discussed by other authors. Moreover, for NPs the terms coalescence and sintering are not strictly differentiated.

In [7] it is noted that coalescence of metal NPs closely relates to the Tammann temperature $T_\mathrm{T} = 0.5 T_\mathrm{m}^{(\infty)}$ where $T_\mathrm{m}^{(\infty)}$ represents the bulk melting point. According to [7], when $T_\mathrm{T}$ is reached, the atomic thermal motion results in the interparticle diffusion and, therefore, in the NP coalescence. However, the NP melting temperature $T_\mathrm{m} < T_\mathrm{m}^{(\infty)}$ seems to be a more adequate criterion of the coalescence/sintering scenario transformation as $T \geq T_\mathrm{m}$ corresponds to the temperature region where NPs do not keep their crystalline structure, i.e. behave themselves as nanosized droplets. So, in our previous papers [8-10] we proposed to interpret the case when $T \geq T_\mathrm{m}$ as the nanodroplet coalescence and the case when $T < T_\mathrm{m}$ as the solid NP sintering. Experimental results [11-13] and our MD results demonstrate different regularities and mechanisms of coalescence and sintering on the nanoscale. Coalescence may be interpreted as a hydrodynamic phenomenon on the nanoscale whereas sintering of two solid NPs reduces to the single or two-grain nanocrystal formation with possible following recrystallization of the two-grain structure i.e. its transformation into a single crystal.

The available experimental studies and MD simulations of the coalescence/sintering on the nanoscale have been performed, first of all, on Au NPs [1-3, 5, 6, 8-13]. However, even for this most popular pattern regularities and mechanisms of the processes in question are not quite clear up now. According to [1], coalescence of two metal NPs proceeds in two stages: (i) maximizing the contact area, i.e. the neck formation between the two NPs (Fig. 1a); (ii) sphericization of the daughter NP (Fig. 1b). MD experiments demonstrate that the neck formation is a very fast process taking of order of 100 ps. At the later stage of coalescence, the notion of the neck radius $x_n$ becomes rather conditional (Fig. 1b), i.e. the small semi-axis of an ellipsoid, modeling the shape of the daughter droplet, is usually interpreted as the neck radius. Under the assumption that the volume of the system is not changed in the course of coalescence, the final (maximal) reduced neck radius $(x_n/r_0)_\mathrm{max}$, corresponding to the spherical daughter droplet shape, will be equal to 1.26, where $r_0$ is the initial NP radius. However, according to Kobata et al. [14], the reduced neck radius of coalescing spheres should reach a smaller maximal value $(x_n/r_0)_\mathrm{max} = 0.83$ found by employing the capillary induced surface diffusion (CISD) concept [15]. Later the CISD concept was additionally developed by Coblenz et al. [16] considering not only the surface diffusion but also the grain boundary diffusion. In accordance with the CISD concept, the negative curvature of the neck surface results in a negative value of the chemical potential treated as the driving force of the surface diffusion of atoms into the neck region. The CISD theory [18] predicts that a current value $x_n$ of the neck radius is achieved during time

$$t = \frac{(x_n/r_0)^6 r_0^4 RT}{CWD\sigma\Omega} \qquad (1)$$

where $C$ is a numerical constant, $R$ is the molar gas constant, $D$ is the surface diffusion coefficient, $\Omega$ is the molar volume, $W$ is the width of a diffusive layer, and $\sigma$ is the surface tension. If $(x_n/r_0) = (x_n/r_0)_{max}$, time $t$ in formula (1) will be equal to the characteristic coalescence/sintering time

$$\tau = \frac{(x_n/r_0)^6_{max}}{C} \frac{RT}{WD\sigma\Omega} r_0^4. \qquad (2)$$

The last equation predicts that $\tau$ should be proportional to $r_0^4$. However, according to the available results of MD simulations [1, 2, 8-10], coalescence of gold nanodroplets, follows neither dependence (1) nor formula (2). At the same time, some authors [5] believed that their MD experiments confirmed the CISD theory. So, in spite on a long enough history of studying coalescence and sintering on the nanoscale, there remain a number of debatable questions and controversial opinions. For this reason, we have been trying to elucidate the regularities and mechanisms of these processes. In particular, we will try to justify our hypothesis that regularities and mechanisms of the nanodroplet coalescence are fundamentally different from those of the solid NP sintering.

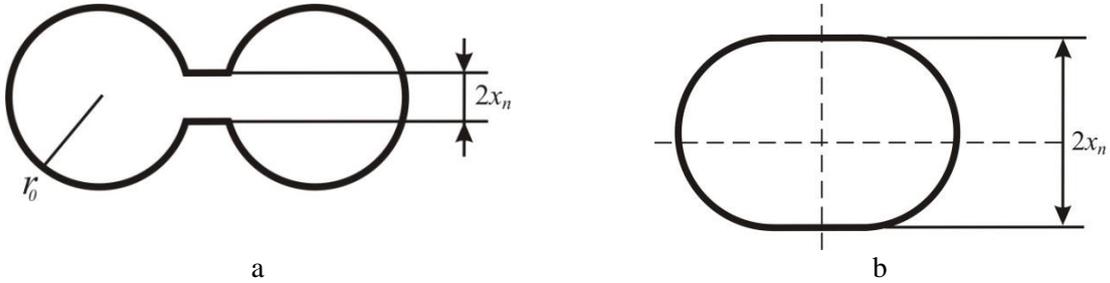

Fig. 1. Schematic presentation of a pair of coalescing nanodroplets at an early (a) and later (b) stages of the process.

In our former works [8-10] we concluded that temperature $T = T_m(N_0)$ corresponds to a transition from the solid NP sintering scenario ($T < T_m$) to that of the nanodroplet coalescence. However, in our MD experiments [8-10] chosen temperatures were higher than $T_m$ by at least 50 K or lower than $T_m$ by at least 50 K, i.e. the region from $T_m - 50$ K to $T_m + 50$ K was not explored in detail. So, it remains unclear whether just $T = T_m$ may be interpreted as a critical (characteristic) temperature of switching from scenario of the solid NP sintering to that of the liquid nanodroplet coalescence. The present paper is devoted to MD investigation of the coalescence/sintering regularities and mechanisms just in the vicinity of the NP melting temperature $T_m$. We have found that the sintering scenario is switched to the coalescence one not at $T = T_m$ exactly by at a lower characteristic (critical) temperature $T_0 \approx 0.9 T_m$. In turn, we have found that the sintering – coalescence transition at $T = T_0$ may be interpreted as a bifurcation phenomenon.

## 2. Approaches to atomistic simulation and processing simulation results

The isothermal MD simulations were performed by using open and well-verified program LAMMPS. This program makes it possible to involve parallel calculations on graphical processing units (GPUs) that significantly extends both the size rang of the simulated objects and the evolution time reproduced in MD experiments. The interatomic interactions were described by employing the embedded atom method (EAM) with parametrization recommended for Au in [17]. The size dependence of the Au NP melting temperature $T_\mathrm{m}$, obtained in our previous MD experiments [9] and compared to the available experimental data is presented in Fig. 2. This dependence will be used to interpret our new MD results on coalescence/sintering of Au NPs. At the same time, agreement between experimental and MD results for $T_\mathrm{m}$ confirms adequacy of parametrization [17] to Au NPs in a wide enough size range. To simulate the coalescence/sintering process, a spherical fragment of the bulk Au lattice was relaxed (annealed), duplicated and then put into a point contact with its copy providing an initial gap of order of the atomic size between NPs.

In the present study we have found and analyzed kinetic dependences for two phenomenological parameters of the coalescence/sintering process. First of them is the reduced neck radius $x_n^* = x_n/r_0$ where $x_n$ is the neck radius (see Fig. 1a) and $r_0$ is the initial NP radius. The second parameter is the shrinkage coefficient $\xi = 1 - L/L_0$, where $L$ is the distance between the centers of mass of the coalescing/sintering NPs and $L_0$ is the initial value of $L$ ($L_0 \approx 2r_0$). Evaluations of these parameters by other authors and previous investigations of their kinetic behavior were discussed in our papers [8-10]. To characterize the structure of coalescing/sintering NPs and of the resulting (daughter) NP, the degree of crystallinity $\eta$ and the radial distribution function $g(r)$ were calculated by employing the Ovito program. The $x_n^*$ and $\xi$ parameters were determined by employing our own program developed for the MD data processing.

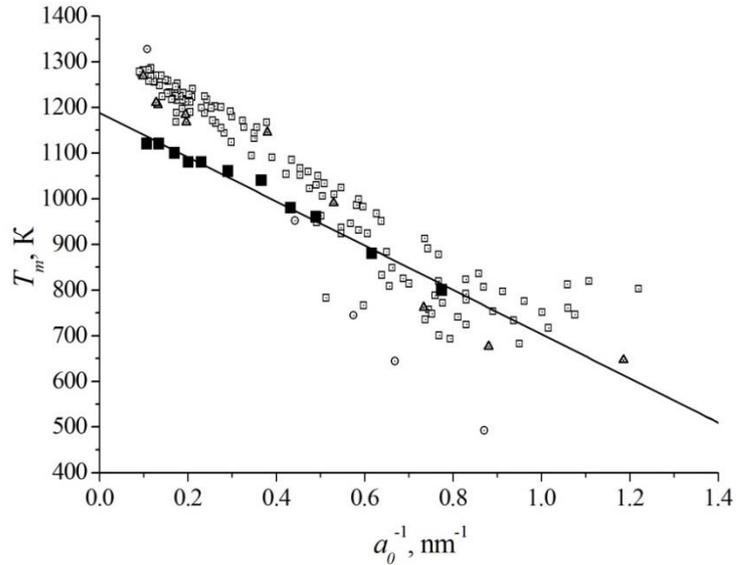

Fig. 2. Size dependence of the melting temperature $T_m$ of Au NPs. Dots ■ and the solid straight line present our MD results. Other (open) symbols correspond to experimental data [18] (□), [19] (○) and [20] (Δ).

## 3. Results and discussion

In order to determine the characteristic temperature $T_0$ for Au NPs, consisting each of 30000 atoms, we have performed three series of MD experiments. First of them corresponds to MD simulations of coalescence/sintering in the temperature range from 1050 K to 1100 K with the temperature increment $\Delta T = 50$ K. In other words, these MD experiments were performed at $T = 1000, 1050$ and $1100$ K. The chosen temperature range contains the melting temperature $T_\mathrm{m} = 1090$ K determined earlier just for NPs of the chosen size (see Fig. 2). The second series of MD experiments corresponds to $\Delta T = 10$ K and the third to $\Delta T = 2$ K. Employing such an approach, we will be able to determine $T_0$ with the accuracy of 2 K.

3.1. *The first series of MD experiments ($\Delta T = 50$ K)*

In Fig. 3 a snapshot is presented of a daughter nanodroplet formed after the coalescence time of 1.1 ns. One can see that for this very short time interval the daughter droplet acquires almost perfect spherical shape. Here and in following figures we color differently atoms belonging to different NPs (yellow and gray colors) to better understand the structural rearrangements in the course of the coalescence and sintering processes. For the same time of 1.1 ns solid NP (Fig. 3b) forms a dumbbell bicrystal, i.e. a two-grain structure with a grain boundary between the grains in the neck region. The neck formation is typical not only for the droplet coalescence at the initial very fast stage of the process but for sintering solid NPs as well. However, contrary to the nanodroplet coalescence, configuration shown in Fig. 1b remains practically unchanged up to the MD evolution time of 100 ns.

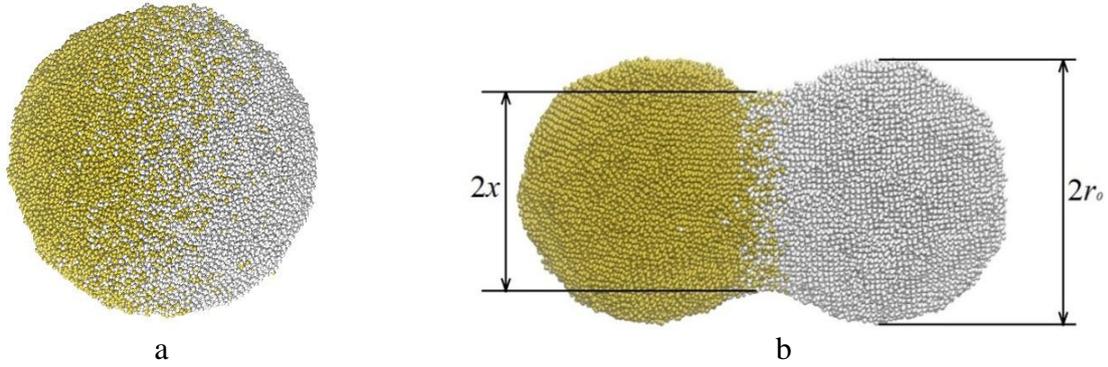

Fig. 3. Snapshots of a MD daughter Au nanodroplet consisting of $N = 2N_0 = 60000$ atoms (a) and of a daughter solid NP (b). Snapshot (a) corresponds to $T = 1100$ K, snapshort (b) to $T = 1000$ K. The melting temperature $T_\mathrm{m}$ of Au NPs, consisting of 30000 atoms is 1090 K [9].

According to Fig. 4, at $T = 1100$ K the reduced neck radius $x_n^*$ very quickly (for about 0.1 ns) reaches its asymptotic value of 1.2 corresponding to the spherical shape of the daughter droplet of the twice greater volume in comparison of that of each of the initial droplets.

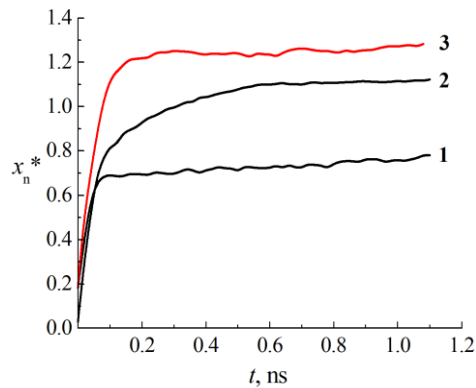

Fig. 4. Kinetic dependences for the reduced neck radius $x_n^*$ corresponding to coalescence/sintering of Au NPs, consisting each of 30000 atoms. Curve 1 corresponds to temperature of 1000 K, curve 2 to 1050 K and curve 3 to 1100 K.

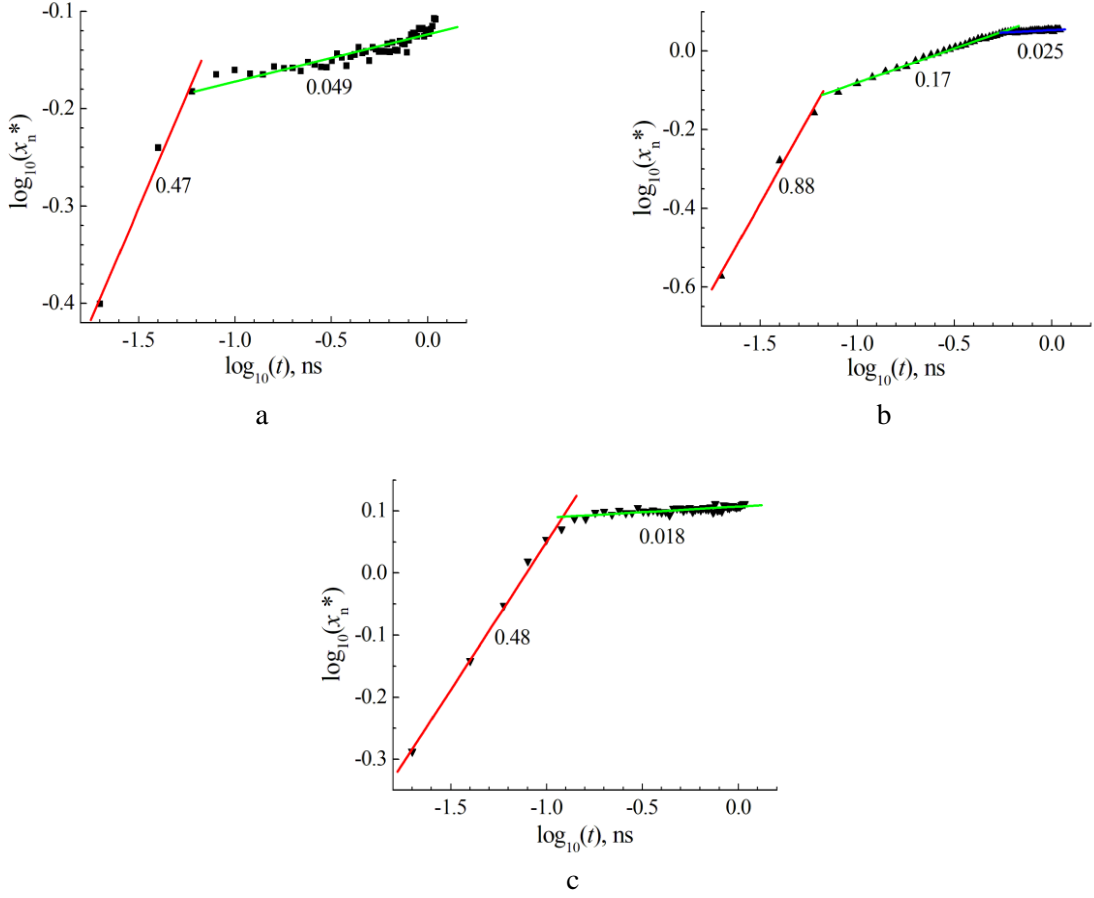

Fig. 5. Dependences of $\log_{10}(x_n^*/r_0)$ on $\log_{10}(t)$ corresponding to coalescence/sintering of two Au NPs consisting each of 30000 atoms at temperatures 1000 K (a), 1050 K (b) and 1100 K (c). Figures under the straight lines correspond to their slope coefficients.

The dependences of $\log_{10}(x_n^*/r_0)$ on $\log_{10}(t)$ are of interest as they make it possible to determine the value of the parameter $m$ in the power dependence $x_n^* \sim t^n$. According to CISD concept (formulas (1) and (2)), $m = 1/6 = 0.17$. This value of $m$ demonstrates Fig. 5b only corresponding to temperature $T = 1050\,K < T_m = 1090\,K$. So, a conclusion can be made than CISD should not be treated as to the main mechanism either of the nanodroplet coalescence or of the solid NPs sintering. It is also noteworthy that the value of the exponent $m$ in the power dependence $x_n^* \sim t^n$ depends on the time interval for which the $\log_{10}(x_n^*/r_0)$ on $\log_{10}(t)$ dependence is plotted. Really, Fig. 6, corresponding to coalescence/sintering times up to 10 ns, demonstrates very low values of $m$: from 0.02 to 0.05.

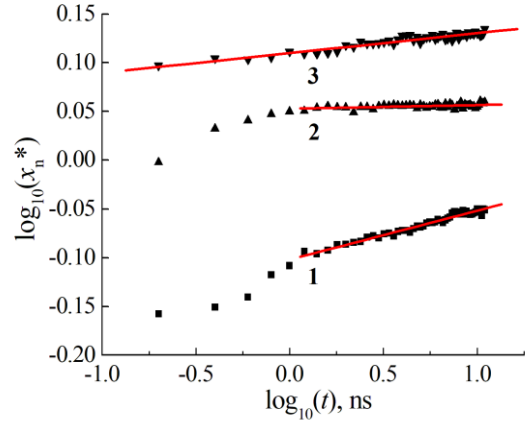

Fig. 6. The $\log_{10}(x_n^*/r_0)$ on $\log_{10}(t)$ dependences for two Au NPs ($N_0 =30000$) coalescing/sintering for 10 ns. Straight line 1 corresponds to $T =1000$ K (the slope coefficient is 0.02), line 2 to 1050 K (the slope coefficient is 0.0037) and line 3 to 1100 K (the slope coefficient is 0.02).

The kinetic dependencies for the shrinkage coefficient are presented in Fig. 7. One can see that curve 3 in Fig. 7b noticeably differs from curves 1 and 2. Really only at $T =1100$ K the $\xi(t)$ dependence approaches the limiting value of $\xi$ equal to unity.

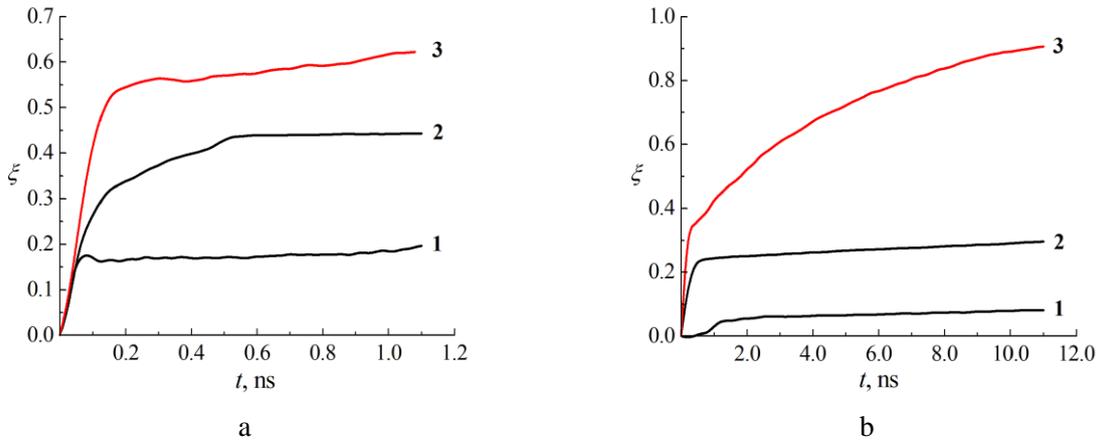

Fig. 7. Kinetic dependences of the shrinkage coefficient (ratio) corresponding to two Au NPs consisting each of 30000 atoms, Fig. 1a corresponds to the time scale of 1 ns, Fig. 1b to 10 ns. Curve 1 represent our MD results for $T =1000$ K, curve 2 MD results for $T =1050$ K and curve 3 MD results for $T =1100$ K.

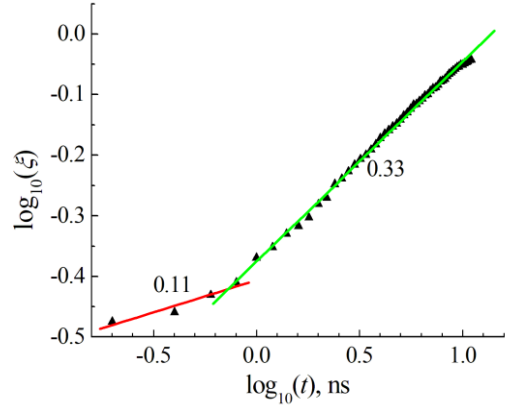

Fig. 8. Dependence of $\log_{10}(\xi)$ on $\log_{10}(t)$ corresponding to $T =1100$ K.

We have also tried to plot the $\log_{10}(\xi)$ on $\log_{10}(t)$ dependence (Fig. 8). One can see that a transition takes place from the linear dependence with the slope coefficient of 0.11 to the linear dependence with the slope coefficient of 0.33. The transition under discussion occurs at $\log_{10}(t) \cong 0$, i.e. $t \cong 1$ ns. This value of $t$ corresponds to the complete sphericization of the daughter nanodroplet. In the other words, the second linear segment ($t > 1$ ns) corresponds to the self-diffusion of atoms inside the spherical daughter nanodroplet.

*3.2. The second series of MD experiments (ΔT =10 K)*

Fig. 9 demonstrates kinetic dependences of the $x_n^*$ and $\xi$ parameters corresponding to for the second series of MD experiments. The main conclusion drawn from these dependences is that they sharply change their behavior not at $T = T_m$ but at a lower temperature of about 1050 K. Really, for higher temperatures parameter $x_n^*$ reaches the asymptotic value of 1.2 corresponding to the spherical daughter NP shape. According to Fig. 10, for temperatures exciding the melting temperature of the initial NPs, the lines corresponding to all the above temperatures practically coincide.

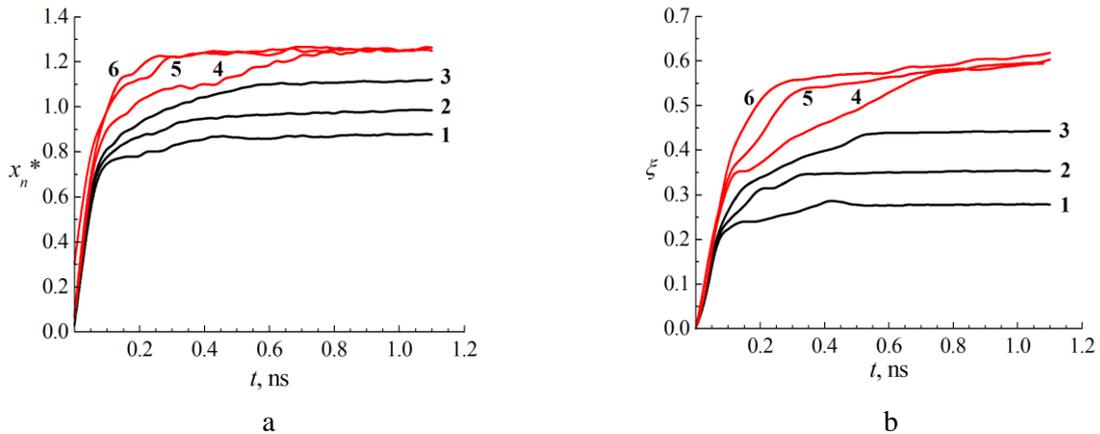

Fig. 9. Kinetic dependences for $x_n^*$ (a) and the shrinkage coefficient $\xi$ (b) corresponding to the second series of MD experiments at temperatures 1030 K (curves 1), 1040 K (curves 2), 1050 K (curves 3), 1060 K (curves 4), 1070 K (curves 5) and 1080 K (curves 6).

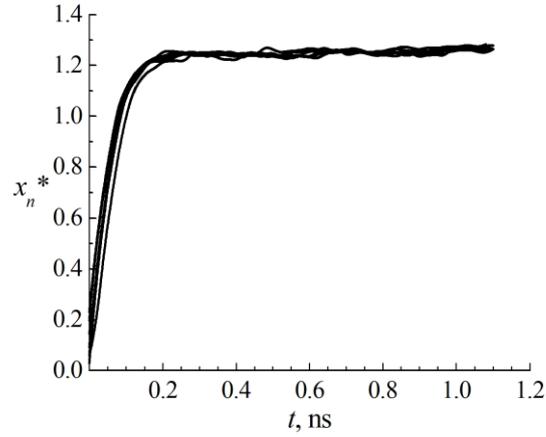

Fig. 10. Kinetic dependences for the reduced neck radius corresponding to temperatures 1090, 1100, 1110, 1120, 1130, 1140, and 1050 K. The lines corresponding to all the above temperatures practically coincide.

*3.3. The third series of MD experiments ($\Delta T = 2$ K)*

The results obtained involving the temperature increment $\Delta T = 2$ K (Fig. 11) seem to be most interesting: the behavior of the $x^*(t)$ and $\xi(t)$ dependences become irregular. For example, usually the shrinkage coefficient $\xi(t)$ grows under growing temperature. However, Fig. 11b demonstrates an alternative behavior of the $\xi(t)$ dependence when a higher value of this parameter (curve 1) corresponds to a lower temperature (1052 K) in comparison with dots ▲ corresponding to a higher temperature of 1054 K. In Fig. 12 kinetic dependences for the shrinkage coefficient are presented for the systems distinguished in the microstates only. In other words, locations and velocities of atoms inside the initial NPs were a bit different from each other but temperature, size and all the other phenomenological parameters were the same.

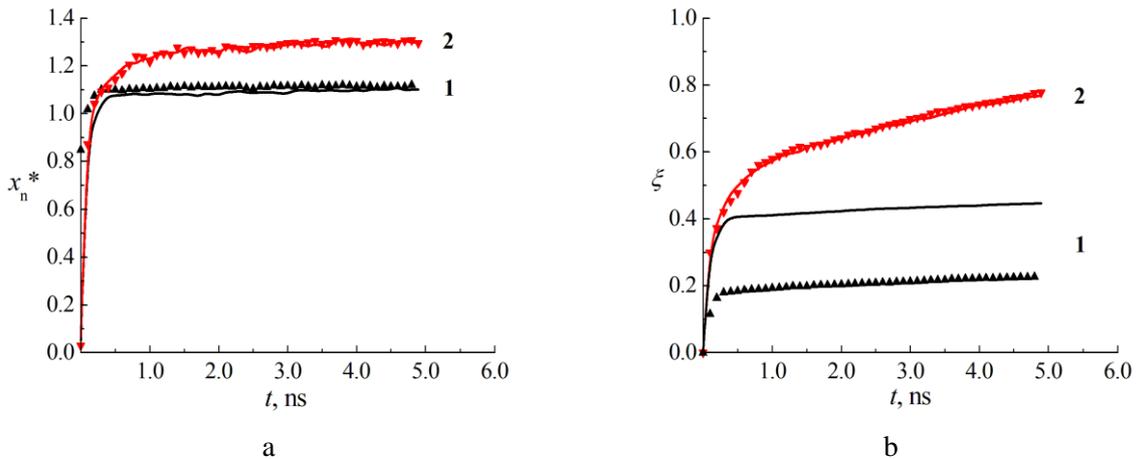

Fig. 11. Kinetic dependences for $x_n^*$ (a) and $\xi$ (b) corresponding to temperatures 1052 K (curves 1), 1054 K (dots ▲), 1056 K (curves 2), 1058 K (dots ▼).

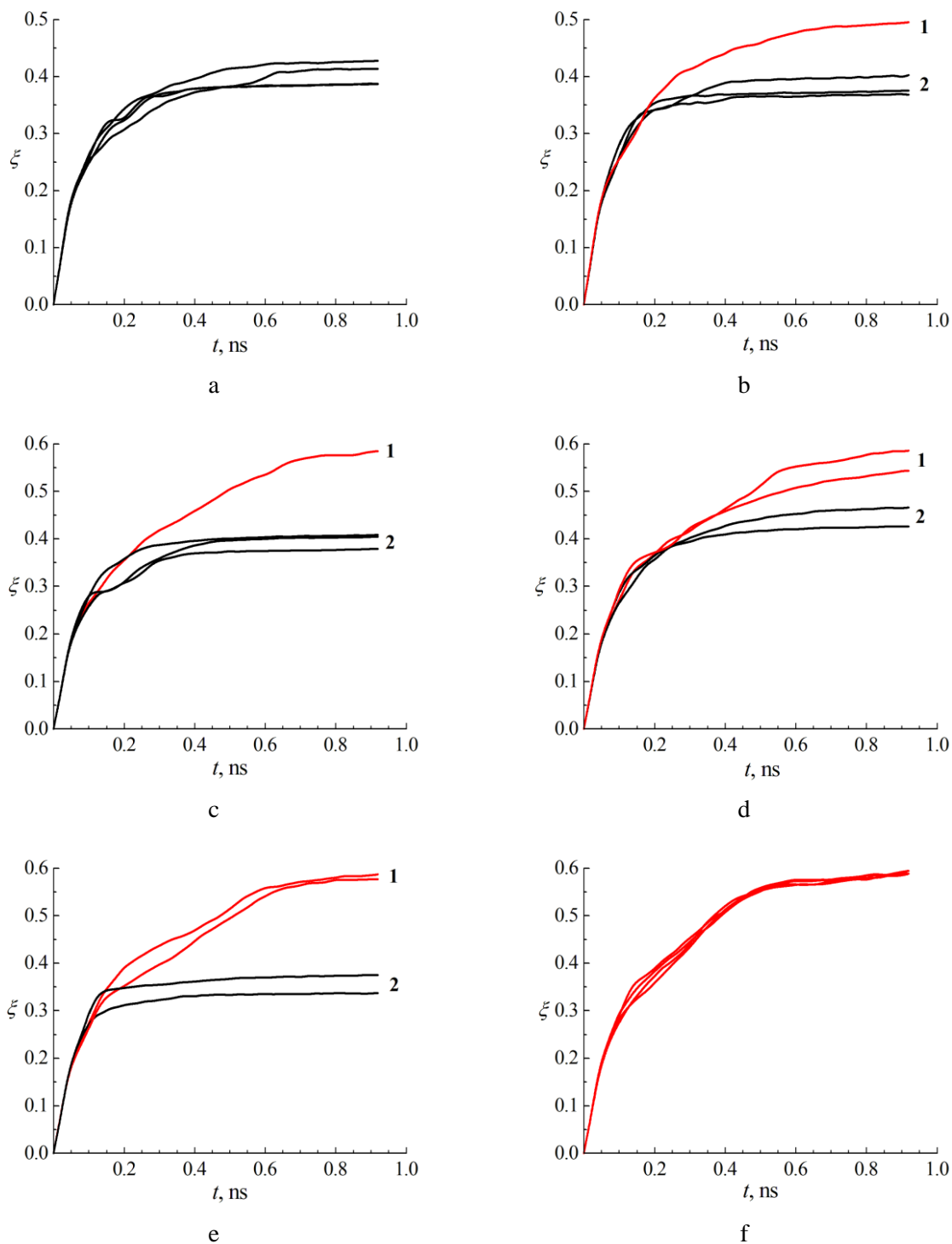

Fig. 12. Kinetic dependences for the shrinkage coefficient $\xi$ corresponding to different temperatures: 1050 K (a), 1052 K (b), 1054 K (c), 1056 K (d), 1058 K (e) 1060 K (f). Red lines 1 correspond to the nanodroplet coalescence, black solid lines to the solid NP sintering.

To elucidate the nature of such an anomalous behavior, we investigated kinetic dependences if the degree of crystallinity $\eta$ using for this purpose the Ovito program. According to Fig. 13, for the same temperature of 1054 K and the same NP size ($N_0 = 30000$ atoms) coalescence/sintering of NPs can follow to two different scenarios. Really, according to Fig. 13a, the first of these scenarios corresponds to $\eta \to 0$ i.e. $(1-\eta) \to 1$ for $t > 0.7$ ns. The second scenarios (Fig. 13b) corresponds to the asymptotic values of about 0.2 and 0.7, respectively. For $T = 1056$ K the $\eta(t)$ dependences are presented in Fig. 14.

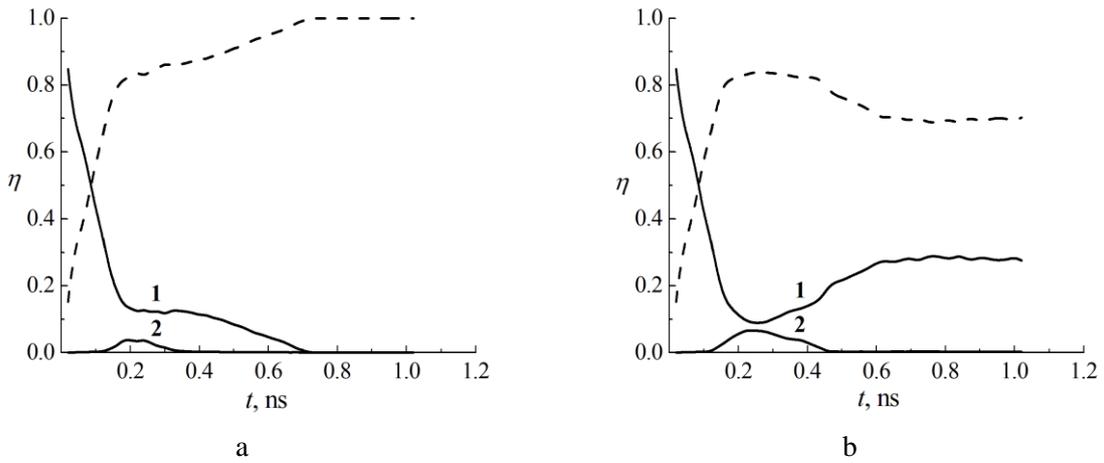

Fig. 13. Two different scenarios (a and b) of the kinetic behavior of the degree of crystallinity $\eta$ at 1054 K. Curve 1 corresponds to the term of the fcc local structure, curve 2 to the term of the bcc structure. Dashed lines correspond to $(1-\eta)$ where $\eta = \eta_{fcc} + \eta_{bcc}$.

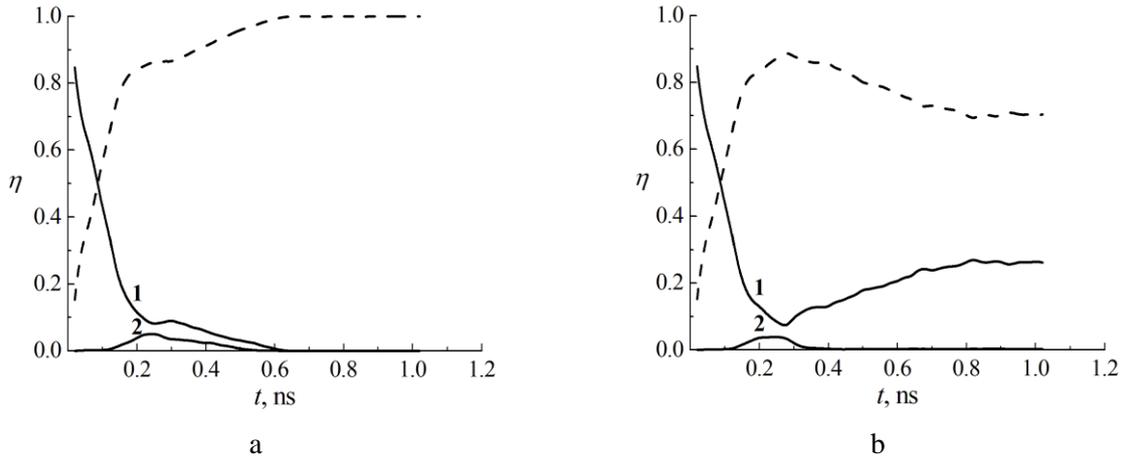

Fig. 14. Two different scenarios (a and b) of the kinetic behavior of the degree of crystallinity $\eta$ at 1056 K. Curve 1 corresponds to the term of the fcc local structure, curve 2 to the term of the bcc structure. Dashed lines correspond to $(1-\eta)$ where $\eta = \eta_{fcc} + \eta_{bcc}$.

According to Fig. 15, in the case of the nanodroplet coalescence, at 0.7 ns the radial distribution function (RDF) (left column) looks like RDF for bulk liquid. The second scenario (scenario of sintering) RDF corresponds rather to the amorphous or crystalline state.

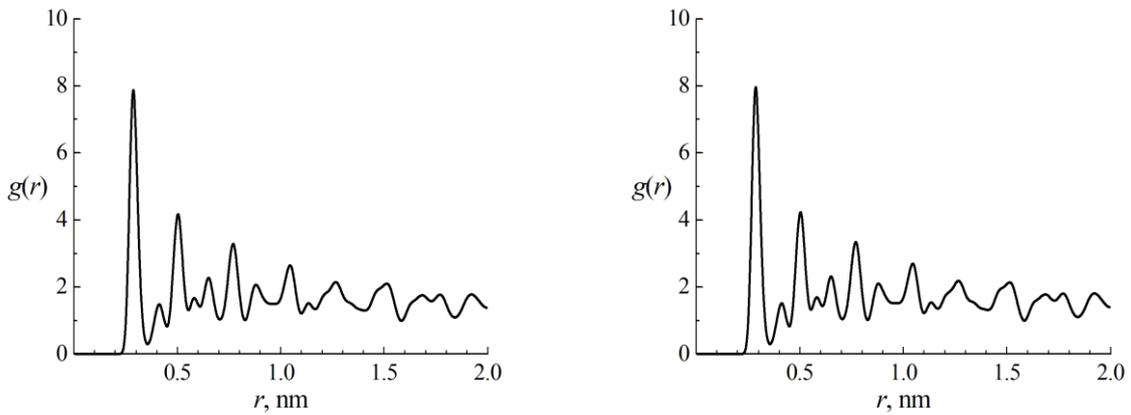

0.02 ns

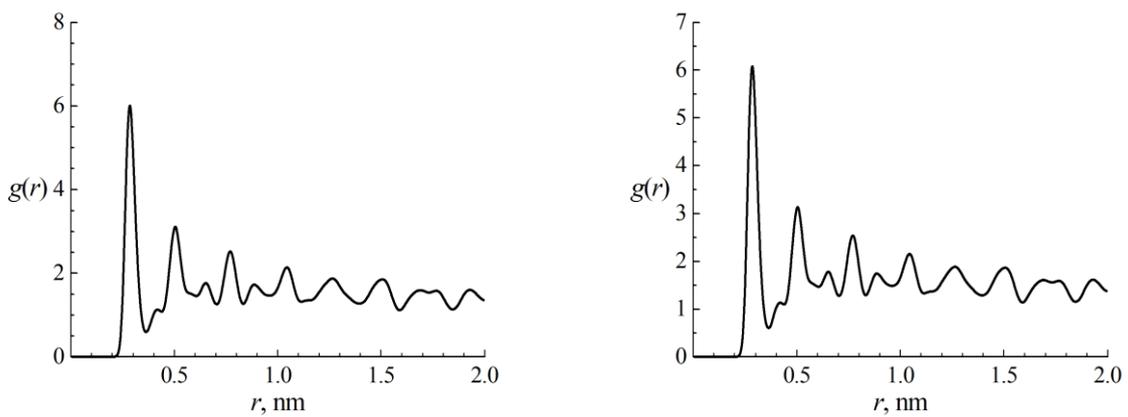

0.1 ns

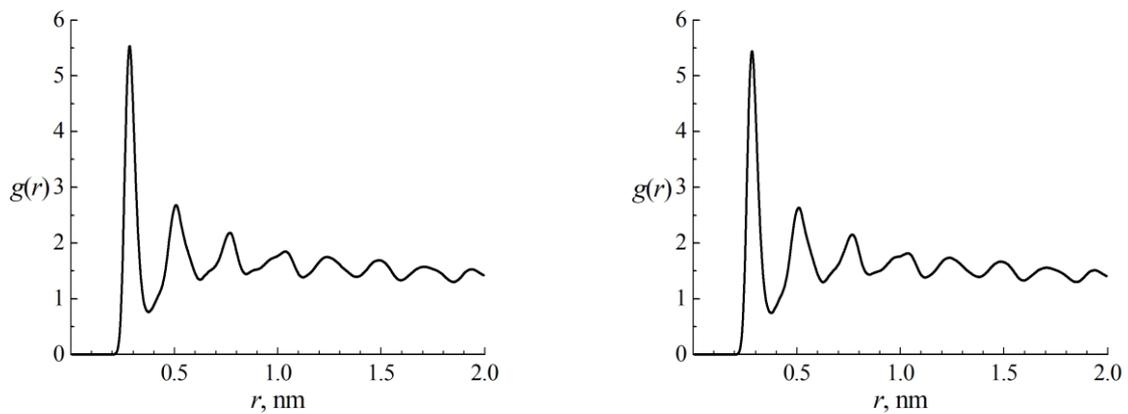

0.16 ns

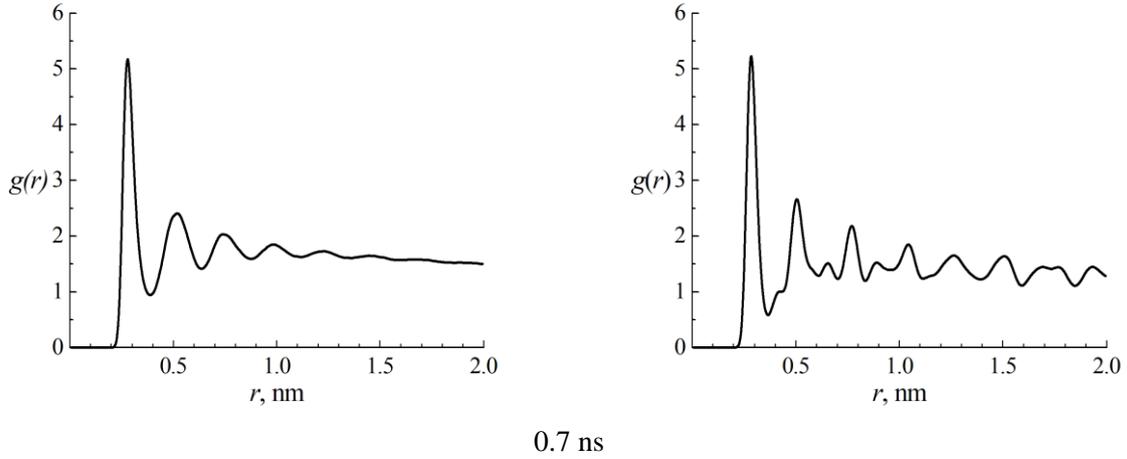

0.7 ns

Fig. 15. Radial distribution functions for two different scenarios (left and right columns) of coalescence/sintering of two Au NPs at $T = 1054$ K ($N_0 = 30000$).

So, in some cases interaction between two solid NPs results in a melted daughter NP, i.e. the process under discussion looks like the droplet coalescence. However, for the same temperature and the same NP size the daughter NP can inherits, to a greater or lesser extent, the crystalline structure of the initial solid NPs. So, it may be assumed that $T_0 = (1055 \pm 1)$ K for Au NPs of the chosen size, i.e. $T_0 \approx 0.97\, T_m$.

## 4. Conclusion

The above MD results confirm our hypothesis that the solid NP sintering scenario is transformed into the coalescence one corresponding to the complete merging of NPs and formation of a daughter nanodroplet occurs not at $T = T_m(N_0)$ exactly but at the characteristic (critical) temperature $T_0 < T_m(N_0)$. For Au NPs, containing each $N_0 = 30000$ atoms, $T_0 = 1055$ K whereas $T_m(30000) = 1091$ K, i.e. the reduced critical temperature $T_0^* = T_0/T_m(N_0)$ is equal to 0.97. We have found that $T = T_0$ corresponds to a bifurcation point when coalescence/sintering can result in formation of both a dumbbell two-grain nanocrystal and a daughter spherical nanodroplet. The probabilities of these scenarios have not been exactly determined yet, but it seems that they are of order of 50 %. As $T_m(2N_0) > T_m(N_0) > T_0$, the daughter nanodroplet should finally transfer into a nanocrystal. However, the characteristic time of such a crystallization should be long enough as $T_0 = 1055$ K is lower than $T_m(2N_0) = 1110$ K by 55 K only.

The above results have been obtained for coalescing/sintering of NPs consisting each of 30000 atoms. In the present paper we have not studied in detail the effect of size on the coalescence/sintering in the vicinity of the melting point. However, in general the same regularities were observed in the case when $N_0 = 10000$ atoms ($N = 20000$ atoms). In particular, we have found that the bifurcation point corresponds to $T_0 = 1047$ K $< T_m(10000) = 1076$ K. It is also noteworthy that the value $T_0 = 0.97$ of the reduced critical temperature is not changed when $N_0$ becomes three times lower in comparison with $N_0 = 30000$.

**Acknowledgments**

The work was supported by the Ministry of Science and Higher Education of the Russian Federation in the framework of the State Program in the Field of the Research Activity (project no. 0817-2020-0007).